\documentstyle[pra,aps,eqsecnum,multicol]{revtex}

\begin{document}
\title{Analysis of Generalized Grover's Quantum Search Algorithms \\
Using Recursion Equations  
}
\author{Eli Biham$^1$, Ofer Biham$^2$, 
David Biron$^2$\footnote{Present address: Department of Physics of
Complex Systems, Weizmann Institute of Science, Rehovot 76100, Israel}, 
Markus Grassl$^3$, 
Daniel A. Lidar$^4$\footnote{Permanent address: Chemistry Department, 
University of Toronto, 
80 St. George Street, Toronto, ON, Canada M5S 3H6} 
and Daniel Shapira$^2$}
\address{
$^1$Computer Science Department, Technion, Haifa 32000, Israel \\
$^2$Racah Institute of Physics, The Hebrew University, Jerusalem
91904, Israel\\ 
$^3$Institut f\"{u}r Algorithmen und Kognitive Systeme,
Universit\"{a}t Karlsruhe, 
\hbox{Am Fasanengarten 5}, D--76128 Karlsruhe, Germany\\
$^4$Department of Chemistry, University of California, Berkeley, CA
94720, USA}

\maketitle
\def\bra#1{\left<#1\right|}
\def\ket#1{\left|#1\right>}

\begin{abstract}
The recursion equation analysis of Grover's quantum search 
algorithm presented by Biham et al. [PRA 60, 2742 (1999)]
is generalized. It is applied to the 
large class of Grover's type algorithms in which the Hadamard 
transform is replaced by any other unitary transformation 
and the phase inversion is replaced by a rotation by an arbitrary
angle.
The time evolution of the amplitudes of the marked and 
unmarked states, 
for any initial complex amplitude distribution 
is expressed using 
first order linear difference equations. 
These equations are solved {\it exactly}. 
The solution provides the number of iterations $T$ after which 
the probability of finding a marked state upon measurement is the highest, 
as well as the value of this probability, $P_{\max}$.  
Both $T$ and $P_{\max}$ 
are found to depend on the averages
and variances of the initial amplitude distributions 
of the marked and unmarked states,
but not on higher moments.
\end{abstract}


\begin{multicols}{2}

\section{Introduction}
Grover's search algorithm 
\cite{Grover96,Grover97a} 
provides a 
dramatic example of the potential speed-up offered by quantum computers. 
It also provides an excellent 
laboratory for the analysis and implementation of quantum algorithms in 
various hardware media. 
The problem addressed by Grover's algorithm can be viewed as trying to 
find a marked element in an unsorted database of size $N$. To solve this 
problem, a classical computer would need, on average, 
$N/2$ 
data base queries and $N$ queries in the worst case.  
Using Grover's algorithm, a quantum computer can
accomplish the same task using merely $O(\sqrt{N})$ queries. The
importance of Grover's 
result stems from the fact that it proves the enhanced power of
quantum computers compared to classical ones for a whole class of
oracle-based problems, for which the bound 
on the efficiency of classical
algorithms is known.

Grover's algorithm can be represented as searching a
pre-image of an oracle-computable boolean function, 
which can only be computed
forward, but whose inverse cannot be directly computed.
Such a function is $F:  D  \rightarrow \{0,1\}$
where $D$ is a set of $N$ domain values (or states)
and the pre-images of the value 1 are called the marked states.
The problem is to identify one of the marked states, i.e., some
$v \in D$ such that $F(v)=1$.
Problems of this type are very common.
One important example, from cryptography,
is searching for the key $K$ of the Data Encryption Standard
(DES) \cite{Stinson95}, given a known plaintext $P$ 
and its ciphertext $C$,
where $F=1$ if the pair of plaintext and ciphertext match [i.e.,
$E_K(P)=C$ where $E_K$ is the encryption function] and $F=0$ otherwise.
Other examples are solutions of 
nondeterministic polynomial time
(NP) and NP-complete problems, which
include virtually all the difficult computing problems in practice
\cite{Garey:book}.

Let us assume, for simplicity that $N = 2^n$, where $n$ is an integer. 
We introduce a register 
$\left| \bar{x} \right\rangle  = \left| x_{1} \ldots x_{n} \right\rangle$  
of $n$ qubits to be used in the computation. 
Grover's original quantum search algorithm consists of the following steps:
\begin{enumerate}
\item Initialize the register to 
      $\left| \bar{0} \right\rangle = \left| 0 \ldots 00 \right\rangle$,
      and apply the Hadamard transform to obtain 
      a uniform amplitude distribution. 
\label{init:item}
\item  Repeat the following operation $T$ times:

\begin{description}
\item[A.]  Rotate the marked states by a phase of $\pi$ radians.

\item[B.]  Rotate all states by $\pi $ radians around the average
amplitude of {\it all} states. This is done by 
(i) Hadamard transforming every qubit; (ii) rotating the  
 $\left| \bar{0} \right\rangle$ state by a phase of $\pi$ radians, and 
(iii) again Hadamard transforming every qubit. 
\end{description}

\item  Measure the resulting state.
\end{enumerate}

A large number of results followed Grover's discovery. 
These results include the proof 
\cite{Zalka99} 
that the algorithm is as 
efficient as theoretically possible \cite{Bennett97}.   
A variety of applications were developed, in which the
algorithm is used in the solution of other
problems
\cite{Durr96,Grover97b,Grover97c,Terhal98,Brassard98,Cerf00,Grover00,Carlini99}.
Recently, experimental implementations 
of Grover's algorithm
were constructed using 
nuclear magnetic resonance 
(NMR)
\cite{Chuang98b,Jones98}
as well as an optical device
\cite{Kwiat00}. 

Several generalizations of Grover's original
algorithm have been developed. The case in which there are several marked 
states was studied in Ref. \cite{Boyer96}. 
Let $k(t)$ [$l(t)$] 
denote the amplitude of the {\it marked} [{\it
unmarked }] states  
after $t$ iterations of the algorithm.
It was shown in \cite{Boyer96}
that the amplitude of the marked 
states increases as: $k(t)=\sin [\omega (t+1/2)]/\sqrt{r}$,
while at the same time that of the unmarked states decreases as:
$l(t)=\cos [\omega (t+1/2)]/\sqrt{ N-r}$, 
where  
$\omega =2\arcsin (\sqrt{r/N})$ and $r$ is the number of marked states. 
For $N \gg r$ the
optimal time to measure and complete the calculation is after
$T=O(\sqrt{N/r})$ iterations, when $k(t)$ is maximal. 

Recently, the algorithm was further generalized 
by allowing an arbitrary (but constant) unitary
transformation to take the place of the Hadamard transform in the
original setting 
\cite{Grover98a,Gingrich00,Jozsa99}
and an arbitrary phase rotation 
of the marked and the predefined states 
instead of the $\pi$ inversion
\cite{Long99a}. 
Another generalization was obtained 
by allowing for an {\em arbitrary complex initial amplitude
distribution},
instead of the 
uniform initial amplitude distribution
obtained in step 1 above
\cite{Biron97,BihamE99}. 

In this paper we analyze the time-evolution of the amplitudes
in the most general algorithm, 
using an arbitrary unitary transformation and phase rotations
on an arbitrary complex initial amplitude distribution. 
Using first order linear difference equations 
we obtain an exact solution for the time evolution 
of the amplitudes under the most general conditions. 
The solution provides the optimal number of iterations $T$ after which 
the probability of finding a marked state upon measurement is the highest, 
as well as the value of this probability, $P_{\max}$.  
Both $T$ and $P_{\max}$ 
are found to depend on the averages
and variances of the initial amplitude distributions 
of the marked and unmarked states,
but not on higher moments.

The paper is organized as follows. 
In Sec. II we present the generalized algorithm. 
The analysis based on recursion equations is given  
in Sec. III.
The results are discussed in Sec. IV and summarized in Sec. V.
In the Appendix we obtain upper bounds on some weighted averages 
of the initial amplitudes
of the marked and unmarked states
which are needed in the analysis. 

\section{The Generalized Grover Algorithm}

In the generalized algorithm 
the initialization step is
modified as 
follows. 
Instead of initializing the register
according to step 1, 
any initial 
distribution of marked and unmarked states 
can be used
(e.g. the final state of 
any other quantum computation). 
In addition, the $\pi$ phase rotation of marked states described
by $I_f^{\pi} = \sum_{x}e^{i \pi F(x)} \ket{x}\bra{x}$ is replaced by 
rotation by an arbitrary phase 
$\gamma$:
\begin{equation}
I_f^{\gamma} = \sum_{x}e^{i \gamma F(x)} \ket{x}\bra{x}.
\end{equation}
\noindent
The rotation about the average, 
described by $G=- W I_{0}^{\pi} W^{\dagger}$,
where $W$ is the Hadamard transform and 
$I_{0}^{\pi} = I - 2 \ket{0} \bra{0}$, is modified in two ways. 
First, the rotation of 
$\left| 0 \right\rangle$ 
by $\pi$ is replaced by 
the rotation of a predefined basis state
$\ket{s}$ by an angle $\beta$.
Second, the Hadamard transform $W$ is
replaced by an arbitrary unitary operator $U$.
In this generalized algorithm,
step 2B above is replaced by
\begin{equation}
G = - U I_{s}^{\beta} U^{\dagger} 
\label{eq:G}
\end{equation}
\noindent
where 
\begin{equation}
I_{s}^{\beta} = I-(1-e^{i\beta})\ket{s}\bra{s}.
\label{eq:Itaubeta}
\end{equation}

\noindent
In the generalized algorithm, the geometric interpretation of step
2B as a rotation around the average amplitude of all states 
is not straightforward. We will demonstrate below that, in fact, one
can return to this interpretation by identifying suitable variables.
By inserting $I_{s}^{\beta}$
from Eq.
(\ref{eq:Itaubeta})
into
Eq. 
(\ref{eq:G})
we obtain that

\begin{equation}
G = (1-e^{i\beta})\ket{\eta}\bra{\eta}-I
\end{equation}

\noindent
where
$\ket{\eta}= U \ket{s}$.

\section{The Recursion Equations}

\subsection{Analysis}

We will now analyze the time evolution of the amplitudes in the
generalized algorithm with a total of $N$ states,
$r$ of which are marked. 
Let the marked
amplitudes at time $t$ be denoted by 
$k_{i}(t)$, $i\in M$ 
and
the unmarked amplitudes by 
$l_{i}(t)$, $i \in \overline{M}$
where $M$ is the set of marked states ($|M| = r$)
and $\overline{M}$ is the set of unmarked states ($|\overline{M}| = N-r$).
The initial amplitudes,  
$k_{i}(0)$, $i\in M$
and 
$l_{i}(0)$, $i \in \overline{M}$,
at $t=0$ are arbitrary. 
Without loss of generality
we assume that the number of marked states satisfies $1\leq r\leq
N/2$. 
A general state of the system at time $t$ will now take the form

\begin{equation}
\ket{g(t)}=\sum_{i \in M}k_i(t)\ket{i}+\sum_{i
\in\overline{M}}l_i(t)\ket{i}.
\label{ket_gt}
\end{equation}

\noindent
A single Grover iteration 
$G I_f^{\gamma}$
will transform the amplitudes
$k_j(t)$, $j\in M$  
to
$k_j(t+1)=\bra{j}GI_f^{\gamma}\ket{g(t)}$
and the amplitudes
$l_{j}(0)$, $j \in \overline{M}$,
to
$l_j(t+1)=\bra{j}GI_f^{\gamma}\ket{g(t)}$.
We find that the recursion equations describing such iteration
take the form 

\begin{eqnarray}
k_j(t+1) & = &
     (1-e^{i\beta})e^{i\gamma}\eta_j\sum_{i\in M}
                               k_i(t)\eta_i^*
+(1-e^{i\beta})\eta_j \times \nonumber \\
&& \sum_{i\in\overline{M}}
                       l_i(t)\eta_i^*-e^{i\gamma}k_j(t) \\
l_j(t+1) & = &
     (1-e^{i\beta})e^{i\gamma}\eta_j\sum_{i\in M}k_i(t)\eta_i^*
+(1-e^{i\beta})\eta_j \times \nonumber \\
&& \sum_{i\in\overline{M}}l_i(t)\eta_i^*-l_j(t)
\end{eqnarray}

\noindent
where

\begin{equation}
\eta_i = \left\langle i \right. \ket{\eta}, \ \ \ \  i=1,\dots,N.
\end{equation}

\noindent
If $\eta_i=0$ for some $i$, 
the Grover iteration $G I_f^\gamma$
changes only the phase of the state $|i\rangle$. 
Thus the probability to measure a state $|i\rangle$ 
with $\eta_i=0$ remains constant. Hence we can
treat the states with $\eta_i=0$ separately from those with
$\eta_i\ne 0$. 
>From now on
we assume,
without loss of generality,
that for the given operator $U$, the predefined state
$\ket{s}$
is chosen such that
$\ket{\eta}$
satisfies
$\eta_i\ne 0$, $i=1,\dots,N$.
We will now introduce new variables:

\begin{eqnarray}
\label{vars_k:eq}
k'_j(t) &=& \frac{k_j(t)}{\eta_j} \\
\label{vars_l:eq}
l'_j(t) &=& \frac{l_j(t)}{\eta_j}.
\end{eqnarray}
\noindent
With these variables, 
the recursion equations will take the 
form

\begin{eqnarray}
k'_j(t+1)&=&
(1-e^{i\beta})e^{i\gamma}\sum_{i\in M}k'_i(t)|\eta_i|^2
+(1-e^{i\beta}) \times \nonumber \\
&& \sum_{i\in\overline{M}}l'_i(t)|\eta_i|^2-e^{i\gamma}k'_j(t)
\\
l'_j(t+1)&=&
 (1-e^{i\beta})e^{i\gamma}\sum_{i\in M}k'_i(t)|\eta_i|^2
+(1-e^{i\beta}) \times \nonumber \\
&& \sum_{i\in\overline{M}}l'_i(t)|\eta_i|^2-l'_j(t).
\end{eqnarray}

\noindent
Let us define:

\begin{equation}
2C(t) = (1-e^{i\beta})e^{i\gamma}\sum_{i\in M}k'_i(t)|\eta_i|^2
      +(1-e^{i\beta})\sum_{i\in\overline{M}}l'_i(t)|\eta_i|^2.
\label{eq:C}
\end{equation}

\noindent
It becomes clear that 
the time evolution of all the amplitudes 
(of both marked and unmarked states)  
can be expressed by

\begin{eqnarray}
k'_j(t+1)   &=& 2C(t)-e^{i\gamma}k'_j(t)\label{eq_k}, \ \ \ \  {j\in M} \label{eq:recur_k}\\
l'_j(t+1)&=& 2C(t)-l'_j(t), \ \ \ \ \ \  {j\in\overline{M}} {\rm .}
\label{eq:recur_l}
\end{eqnarray}

\noindent
One can define the following weights
\begin{eqnarray}
W_k&=&\sum_{i\in M}|\eta_i|^2 \\
W_l&=&\sum_{i\in\overline{M}}|\eta_i|^2
\label{eq:weights}
\end{eqnarray}

\noindent
quantifying the projections of the state
$\ket{\eta}$
on 
$M$ 
and
$\overline{M}$, 
respectively.
>From the normalization of the state
$\ket{\eta}$
it is clear that
$W_k + W_l = 1$.
These weights can be used in order to define weighted 
averages of the new variables, of the form

\begin{eqnarray}
\bar{k}'(t)    &=&  { {\sum_{i\in M}|\eta_i|^2 k'_i(t)} \over W_k } 
\label{eq:kaverage} \\
\bar{l}'(t) &=&  { {\sum_{i\in\overline{M}}|\eta_i|^2 l'_i(t)} \over W_l}.
\label{eq:laverage}
\end{eqnarray}

\noindent
Using these averages to express $C(t)$ we obtain

\begin{equation}
2C(t) = (1-e^{i\beta}) 
\left[ e^{i\gamma} W_k \bar k'(t) + W_l \bar l'(t) \right].
\label{eq:Csimp}
\end{equation}

\noindent
By averaging over all the marked states in Eq. 
(\ref{eq:recur_k})
and over all the unmarked states in Eq.
(\ref{eq:recur_l})
we find that the weighted averages 
$\bar{k}'(t)$
and 
$\bar{l}'(t)$ 
obey the following recursion equations

\begin{eqnarray}
\bar{k'}(t+1) &=&2 C(t)-e^{i\gamma}\bar{k'}(t)         \label{eq:k-ave} \\
\bar{l'}(t+1) &=&2 C(t)-\bar{l'}(t){\rm .}  
\label{eq:l-ave}
\end{eqnarray}

\noindent
These equations can be solved for 
$\bar{k'}(t)$ 
and 
$\bar{l'}(t)$, 
and along with the initial distribution this 
yields the exact solution for
the dynamics of all amplitudes.
We will proceed to solve the recursion formulae for arbitrary complex
initial amplitudes. 
Let us rewrite Eqs.~(\ref{eq:k-ave})~and
(\ref{eq:l-ave})
in a matrix notation 
\begin{equation}
\vec{r}(t+1) = A \vec{r}(t) 
\label{eq:rt1Mrt}
\end{equation}
where 
\begin{equation}
\vec{r}{(t)} = 
\left( \begin{array}{c} 
       \bar{k}'(t) 
       \\ 
       \bar{l}'(t) 
       \end{array}  \right) 
\label{eq:r}
\end{equation}
and 
\begin{equation}
A = 
\left( \begin{array}{cc} 
       a & b \\
       c & d
      \end{array}  \right). 
\label{eq:M}
\end{equation}

\noindent
The matrix elements of $A$ are given by:
\begin{eqnarray}
a &=& (1 - e^{i\beta})e^{i\gamma}W_k - e^{i\gamma}  \label{eq:a} \\ 
b &=& (1 - e^{i\beta})W_l  \label{eq:b} \\ 
c &=& (1 - e^{i\beta})e^{i\gamma}W_k   \label{eq:c} \\ 
d &=& (1 - e^{i\beta})W_l - 1.  
\label{eq:d} 
\end{eqnarray} 

\noindent
The time-evolution of 
$\bar{k}'(t)$
and
$\bar{l}'(t)$
is given by

\begin{equation}
\vec{r} (t) = A^t \vec{r} (0). 
\label{eq:rtMr0}
\end{equation} 

\noindent
In order to obtain explicit expressions for 
$\bar{k'}(t)$ 
and 
$\bar{l'}(t)$ 
we consider the diagonal matrix: 
\begin{equation}
A_D = S^{-1} A S \equiv 
\left( \begin{array}{cc} 
       \lambda_{+} & 0 \\ 
       0 & \lambda_{-}
      \end{array}  \right). 
\label{eq:MD}
\end{equation} 

\noindent
The eigenvalues
$\lambda_{\pm}$ 
of the matrix 
$A$ 
are the solutions of 
$\det(A - \lambda I) = 0$. 
They can be expressed as
\begin{equation}
\lambda_{\pm} = e^{i \omega_{\pm}} 
\label{eq:lambda+/-}
\end{equation}

\noindent
where

\begin{equation}
\omega_{\pm} = \pi + \frac{\beta + \gamma}{2} \pm \omega 
\label{eq:omega+/-}
\end{equation}

\noindent
and the angular frequency
$\omega$ is in the range $0 < \omega< \pi$, and satisfies: 
\begin{equation}
\cos \omega = W_k \cos \frac{\beta + \gamma}{2} + 
              W_l \cos \frac{\beta - \gamma}{2}.
\label{eq:cosw}
\end{equation}
The matrix 
$S$,
which consists of  
the corresponding column eigenvectors, 
takes the form

\begin{equation}
S = 
\left( 
\begin{array}{cc} 
1 & 1 \\ 
\frac{\lambda_{+} - a}{b} & \frac{\lambda_{-} - a}{b} 
\end{array} 
\right). 
\label{eq:S_matrix}
\end{equation}

\noindent
We will now apply Eq.~(\ref{eq:MD}) 
to reconstruct the matrix $A$, 
and compute $A^t$.
A simpler expression for the time-evolution is obtained:

\begin{equation}
\vec{r}(t) =  S {A_D}^t S^{-1} \vec{r}(0), 
\label{eq:rtr0}
\end{equation} 

\noindent
where 

\begin{equation}
{A_D}^t = 
\left( \begin{array}{cc} 
       {\lambda_{+}}^t & 0 \\ 
       0 & {\lambda_{-}}^t
      \end{array}  \right). 
\label{eq:MD^t} 
\end{equation} 

\noindent
The time dependence is given by

\begin{eqnarray}
\bar{k}'(t)    &=&   z_{1} e^{i \omega_{+} t} - z_{2} e^{i \omega_{-} t} 
\label{eq:k-explicit_ave} \\
\bar{l}'(t)    &=&   z_{3} e^{i \omega_{+} t} - z_{4} e^{i \omega_{-} t} 
\label{eq:l-explicit_ave} 
\end{eqnarray}

\noindent
where

\begin{eqnarray}
z_1 &=& 
\frac{(\lambda_{-}-a) \bar{k}'(0)-b \bar{l}'(0)}{\lambda_{-} - \lambda_{+}} 
\label{eq:z1} \\  
z_2 &=& 
\frac{(\lambda_{+}-a) \bar{k}'(0)-b \bar{l}'(0)}{\lambda_{-} - \lambda_{+}} 
\label{eq:z2}  
\end{eqnarray} 

\noindent
and 

\begin{eqnarray}
z_3 &=& \frac{\lambda_{+} - a}{b} z_1 \label{eq:z3} \\ 
z_4 &=& \frac{\lambda_{-} - a}{b} z_2 \label{eq:z4}. 
\end{eqnarray} 

\noindent
We have now completed the solution for the time dependence of the averages
$\bar{k}'(t)$
and 
$\bar{l}'(t)$.
However, our aim is to obtain the time evolution of
the individual variables
${k_i}'(t)$
and 
${l_i}'(t)$
and from them to extract the amplitudes
${k_i}(t)$
and 
${l_i}(t)$.
Subtracting Eq.~(\ref{eq:k-ave}) 
from Eq.~(\ref{eq:recur_k}), 
and Eq.~(\ref{eq:l-ave}) 
from Eq.~(\ref{eq:recur_l}) 
one finds that

\begin{eqnarray}
k'_i(t+1)-\bar{k}'(t+1) &=& -e^{i \gamma}[k_i'(t)-\bar{k}'(t)] \\
l'_i(t+1)-\bar{l}'(t+1) &=& -[l'_i(t)-\bar{l}'(t)] {\rm ,}
\end{eqnarray}

\noindent 
namely, the difference, in absolute value, between each of the variables
${k_i}'(t)$, ${l_i}'(t)$
and the averages of the corresponding sets are time independent.
This means that

\begin{eqnarray}
\Delta k'_i &\equiv & k'_i(0)-\bar{k}'(0) \label{eq:deltak} \\
\Delta l'_i &\equiv &  l'_i(0)-\bar{l}'(0) \label{eq:deltal} 
\end{eqnarray}

\noindent 
are {\it constants of motion}.
Thus, the
time dependence of the
variables follows

\begin{eqnarray}
k'_i(t) &=& \bar{k}'(t) + (-1)^{t}e^{i \gamma t}
                          \Delta k'_i \label{eq:ki}\\
l'_i(t) &=& \bar{l}'(t) + (-1)^t \Delta l'_i \label{eq:li}{\rm .}
\end{eqnarray}

\noindent
The time evolution of the amplitudes 
can now be obtained:

\begin{eqnarray}
k_i(t) & = & \eta_i[\bar{k}'(t) + (-1)^{t}e^{i \gamma t}
                          \Delta k'_i] \\
l_i(t) & = & \eta_i[\bar{l}'(t) + (-1)^t \Delta l'_i]. 
\end{eqnarray}

\noindent
In this picture all the marked as well as the unmarked states 
evolve in unison so it is sufficient to
follow the time evolution of the average in each set.
The only feature distinguishing the states from one another
is their initial deviation from the average.

\subsection{Results}

>From Eqs.
(\ref{eq:ki})
and
(\ref{eq:li})
it follows immediately that the weighted variances 

\begin{eqnarray}
\sigma_{k}^{2}(t)&=&\frac{1}{W_k} \sum_{i \in M} 
|\eta_i|^2 |k'_{i}(t)-\bar{k'} (t)|^{2}
\label{eq:sigma_k} \\
\sigma _{l}^{2}(t)&=&\frac{1}{W_l} 
\sum_{i \in \overline{M}} |\eta_i|^2 |l'_{i}(t)-\bar{l'} (t)|^{2}
\label{eq:sigma_l}
\end{eqnarray}

\noindent
are time-independent
and therefore at any time $t$
they can be replaced by
$\sigma_{k}^{2} = \sigma_{k}^{2}(0)$
and
$\sigma _{l}^{2} = \sigma _{l}^{2}(0)$,
respectively.
When a measurement is performed at time $t$,
the probability that a marked state will be obtained is
$P(t)=\sum_{i \in M}|k_{i}(t)|^{2}$. 
Since all the operators used are unitary, the
variables 
${k_i}'(t)$
and 
${l_i}'(t)$
satisfy the normalization condition

\begin{equation}
\sum_{i \in M}|\eta_i|^2|k'_{i}(t)|^{2}+
\sum_{i \in \overline{M}}|\eta_i|^2|l'_{i}(t)|^{2}=1
\end{equation}

\noindent
at all times. 
Using the definitions of 
$k'_i(t)$ 
and 
$l'_i(t)$ 
and their weighted averages 
$\bar{k}'(t)$ 
and 
$\bar{l}'(t)$ 
given in Eqs.~(\ref{vars_k:eq})-(\ref{vars_l:eq})
and
({\ref{eq:kaverage}})-({\ref{eq:laverage}}), 
one can bring Eqs.~(\ref{eq:sigma_k}) 
and (\ref{eq:sigma_l}) to the form: 

\begin{equation}
\sum_{i \in M}|k_{i}(t)|^{2}= W_k \sigma_{k}^{2} + W_k |\bar{k}'(t)|^{2}
\label{eq:sig_abs_k2}
\end{equation} 

\begin{equation}
\sum_{i \in \overline{M}} |l_{i}(t)|^{2}= W_l \sigma_{l}^{2} + 
W_l |\bar{l}'(t)|^{2}.
\label{eq:sig_abs_l2}
\end{equation}

\noindent
The first equation provides the probability to measure a marked state
at time $t$, while the probability to measure an unmarked state is
given by the second equation.
We will now try to examine the probability that a measurement at time
$t$ will yield a marked state, and its dependence on the rotation angles
$\beta$ and $\gamma$, the unitary transformation $U$,
the predefined state
$\ket{s}$
and the averages
and standard deviations of the initial amplitude distributions of the
marked and unmarked states.
Using Eq.~(\ref{eq:k-explicit_ave}) 
with
$z_1 = |z_1| e^{i \phi_{1}}$ 
and 
$z_2 = |z_2| e^{i \phi_{2}}$, 
the probability that a measurement at time $t$ 
will yield a marked state
is given by a sinusoidal function of the form  

\begin{equation}
P(t) = P_{av} - \Delta P \cos [2(\omega t + \phi)]
\label{eq:Pt}
\end{equation}

\noindent
where

\begin{equation}
\Delta P = 2 W_k |z_1| |z_2|
\label{eq:deltaP} 
\end{equation}

\noindent
is the amplitude of the oscillations,
$\pi/\omega$ 
is their period,

\begin{equation}
P_{av} = W_k({|z_1|}^2 +{|z_2|}^2 + {\sigma_k}^2) 
\end{equation}

\noindent
is the average, or reference value of the probability,
and

\begin{equation}
2 \phi =  \phi_{1} - \phi_{2}    
\end{equation}

\noindent
is the phase.
These parameters are found to depend on the unitary operator
$U$ which is used in the algorithm, the predefined state
$\ket{s}$
and the angle $\beta$ by which its phase is rotated as well as
the angle $\gamma$ by which the phases of the marked states
are rotated. 
The dependence on the initial amplitudes enters only through the
weighted averages of the variables 
${k_i}'(0)$
and 
${l_i}'(0)$ 
and their standard deviations.

It is thus observed that $P(t)$ does not depend on
any higher moments of the initial amplitude
distribution.  
This is due to the fact that all the transformations in the quantum algorithm
are linear.
Therefore, $k_i(t)$ and $l_i(t)$ can be expressed as some linear combinations
of $k_i(0)$ and $l_i(0)$. The only nonlinearity appears in the expression of
$P(t)$ as a sum of squares of the amplitudes of the marked states at time $t$. 
Therefore, powers higher than quadratic are excluded in the expression for $P(t)$.
Moreover, $P(t)$ does not depend
on any other linear combinations of the first
and second powers of the initial amplitudes, except for the particular weighted
averages that compose the first and second moments.
This results from the fact that 
Grover's iterations 
maintain a large number of conserved quantities,
particularly the variances $\sigma_k^2$ and $\sigma_l^2$ 
which are constants of motion. 
As a result, the time evolution of the amplitudes can be fully 
described by the time dependence of the averages $\bar k'(t)$ 
and $\bar l'(t)$. 
Due to the linearity of the transformations, $\bar k'(t)$ and $\bar l'(t)$ 
can be expresses as linear combinations of $\bar k'(0)$ and $\bar l'(0)$. 
The probability $P(t)$ of measuring a marked state at time $t$ is given by
Eq.~(\ref{eq:sig_abs_k2})
in which the
first term on the right hand side, which includes the second moment
is a constant of motion. Therefore, no other quadratic forms can appear.
The second term depends on the first moment at time $t$, which is related
to the first moment at t=0 through the recursion equations.
Thus, the dependence of
$P(t)$ on the initial amplitudes is only through the first and second moments of their 
distribution.

\section{Discussion}

In order to examine the performance of the generalized Grover algorithm
we will now evaluate
the highest possible probability
$P_{\max }= P_{av} + \Delta P$
that a measurement will yield a marked state.
We will also evaluate
the optimal number of iterations $T$ after which 
the probability 
$P_{\max }$
is achieved.

The limit of difficult search problems is obtained when
$N \gg r \ge 1$.
This is reflected in the fact that in the original Grover
algorithm the probability of measuring a marked state
immediately after the initialization step is
$W_k = r/N$.
The assumption that $W_k \ll 1$ carries over to the generalized 
case discussed here. It is satisfied in all cases except for very
unlikely choices of the state $\ket{s}$ from which one Grover
iteration
with the operator $U$
is sufficient in order to measure a marked state with
high probability. 
In the analysis below 
$W_k$ 
will be considered as a small parameter.
The highest possible value of the probability to 
measure a marked state is: 

\begin{equation}
P_{\max} =  W_k (|z_1| + |z_2|)^2 + W_k {\sigma_k}^2. 
\label{eq:Pmax}
\end{equation} 

Consider the parameters $z_1$ and $z_2$ that express the dependence of  
$P_{\max}$
on the initial amplitudes
and the phase rotation angles $\beta$ and $\gamma$.
The expressions for
$z_1$ and $z_2$ 
in 
Eqs.~(\ref{eq:z1})~and
(\ref{eq:z2})
include
$(\lambda_{-} - \lambda_{+})$ 
in the denominator.
Using Eq. 
(\ref{eq:lambda+/-})  
it can be written as:

\begin{equation} 
\lambda_{-} - \lambda_{+} = 2i {e}^{i \frac{\beta + \gamma}{2}} \sin \omega. 
\label{eq:lambda-_+}
\end{equation}

\noindent
Expanding $\sin \omega$ in powers of $W_k \ll 1$, 
one finds that in case that  
the angles $\beta$ and $\gamma$ are different,
and the difference between them satisfies $|\beta-\gamma| = O(1)$,
(e.g. in radians):

\begin{equation} 
\lambda_{-} - \lambda_{+} = 2i {e}^{i \frac{\beta + \gamma}{2}} 
\sin \frac{|\beta - \gamma|}{2} + O(W_k), 
\label{eq:lambda-_+1}
\end{equation}

\noindent
namely, the denominator, in absolute value, is typically of order unity.
In case 
$\beta = \gamma$ 

\begin{equation} 
\lambda_{-} - \lambda_{+} = 
4i {e}^{i \beta} 
\sin {\beta \over 2} {W_k}^{\frac{1}{2}} + 
O({W_k}^{\frac{3}{2}}) 
\label{eq:lambda-_+2}
\end{equation}

\noindent
and the denominator is of order 
$\sqrt{W_k} \ll 1$.
In the following we discuss the search problem in these two cases separately.    

\subsection{Different Rotation Angles: $\beta \neq \gamma$}

Here we consider the case when 
$\beta \neq \gamma$
and the difference between them 
is fixed and finite, and satisfies
$|\beta-\gamma| = O(1)$.
In this case, 
using Eqs.~(\ref{eq:a}), (\ref{eq:b}), (\ref{eq:lambda+/-}) 
and the assumption that
$W_k \ll 1$,
the matrix elements can be approximated by

\begin{eqnarray}
 \label{eq:aa}
a &=& -{e}^{i \gamma} + O(W_k)   \\
 \label{eq:bb}
b &=& 1-{e}^{i \beta} + O(W_k).
\label{eq:lambda+}
\end{eqnarray}

\noindent
The eigenvalues are

\begin{eqnarray}
\lambda_{+} &=& -{e}^{i \beta} + O(W_k)  \\ 
\label{eq:lambda-} 
\lambda_{-} &=&  -{e}^{i \gamma} + O(W_k)  
\end{eqnarray}

\noindent
when $\beta > \gamma$
($0 < \beta < 2 \pi$, $0 < \gamma < 2 \pi$),
and

\begin{eqnarray}
\lambda_{+} &=& -{e}^{i \gamma} + O(W_k)  \\ 
\label{eq:lambda--} 
\lambda_{-} &=&  -{e}^{i \beta} + O(W_k)  
\end{eqnarray}

\noindent
when $\gamma > \beta$.
In the Appendix it is shown
that the initial amplitude distribution 
satisfies
$|\bar{k}'(0)| = O(W_k^{-1/2})$ 
and 
$|\bar{l}'(0)| = O(1)$. 
>From Eqs.~(\ref{eq:z1}) and (\ref{eq:z2}) 
we obtain that
$|z_1| = O(1)$
and 
$|z_2| = O({W_k}^{-1/2})$,
in case $\beta > \gamma$,
while in case
$\gamma > \beta$,
$|z_1| = O({W_k}^{-1/2})$
and
$|z_2| = O(1)$.
Therefore, in both cases,
using Eq.~(\ref{eq:deltaP})
one obtains

\begin{equation} 
\Delta P = O(W_k^{1/2})
\end{equation}

\noindent
namely, the amplitude of the oscillations is negligible.
Thus, the probability to measure a marked state after any number of
iterations cannot be significantly larger
than the probability, 
given by Eq.~(\ref{eq:sig_abs_k2}), 
to measure a marked state at 
time $t = 0$.
We conclude that in this case 
the algorithm fails to enhance the probability of measuring a marked state. 
The angular frequency is 
$\omega = |\beta-\gamma|/2 + O(W_k)$.
Clearly, for such a high frequency,
for which the period is of the order of only few steps, 
it is hard to exploit the oscillations 
since measurements can be taken only in discrete times
and are likely to miss the highest point.
The analysis presented above applies as long as
$W_k \ll (\beta - \gamma)^2$. 
The conclusions are in agreement with the results of Refs.
\cite{Long99a,Long99b}.

\subsection{Identical Rotation Angles: $\beta = \gamma$}

In this case the matrix elements can be approximated 
according to

\begin{eqnarray}
a &=& -{e}^{i \beta} + O(W_k) \label{eq:aaa} \\
b &=& 1-{e}^{i \beta} + O(W_k) \label{eq:bbb} \\
\lambda_{\pm} &=& -{e}^{i \beta} \mp i \left[2{e}^{i \beta} 
\sin {\beta \over 2} \right] {W_k}^{\frac{1}{2}} + O(W_k). 
\label{eq:lambda_pm} 
\end{eqnarray} 

\noindent
This gives rise to

\begin{eqnarray} 
z_{1,2} &=& 
\frac{1}{2} {W_k}^{-\frac{1}{2}} \left[ {i \bar{l}'(0) {e}^{i \psi} \pm 
{W_k}^{\frac{1}{2}} \bar{k}'(0)}\right] + O(1)  
\label{eq:z12} 
\end{eqnarray} 

\noindent
where $\psi=(\pi - \beta)/2$. 
Inserting Eq.~(\ref{eq:z12}) into Eq.~(\ref{eq:Pmax}) 
we obtain

\begin{eqnarray} 
P_{\max} &=& \frac{1}{4}
 \left[ |i \bar{l}'(0) {e}^{i \psi} + {W_k}^{\frac{1}{2}} \bar{k}'(0)| + 
  |i \bar{l}'(0) {e}^{i \psi} - {W_k}^{\frac{1}{2}} \bar{k}'(0)|
\right]^2 \nonumber \\
&+& W_k {\sigma_k}^2 + O(W_k). 
\label{eq:Pmaxpsi}
\end{eqnarray} 

\noindent
To simplify this expression we will now use the identity 
$(|a+b| + |a-b|)^2 = 2(|a|^2 + |b|^2 + |a^2 - b^2|)$, 
where $a$ and $b$ are complex numbers. 
We will also replace
$\bar{l}'(0)$ 
by 
${W_l}^{\frac{1}{2}} \bar{l}'(0)$ 
[note that ${W_l}^{\frac{1}{2}} = 1 + O(W_k)$]
and find that

\begin{eqnarray} 
P_{\max} = 1 &-& W_l {\sigma_l}^2 - \frac{1}{2}  W_l |\bar{l}'(0)|^{2}  
- \frac{1}{2} W_k |\bar{k}'(0)|^{2} \nonumber  \\  
&+& \frac{1}{2} \left| W_l |\bar{l}'(0)|^{2} {e}^{2i(\psi + \alpha_l - \alpha_k)} + 
W_k |\bar{k}'(0)|^{2} \right| \nonumber \\
&+& O(W_k) 
\label{eq:Pmaxalpha} 
\end{eqnarray} 

\noindent
where $\bar{k}'(0) = |\bar{k}'(0)| {e}^{i \alpha_k}$ 
and 
$\bar{l}'(0) = |\bar{l}'(0)| {e}^{i \alpha_l}$. 
This is in agreement with the results of Ref.
\cite{Gingrich00},
where the case $\beta=\gamma=\pi$
was studied using a different approach.
We observe that
$P_{\max}$ 
depends on the statistical properties 
(averages and variances) of the 
initial amplitude distribution 
of the marked and unmarked states. 
For a given distribution,
the 
probability of measuring a marked state is bounded by 
$P_{\max} =  1 - W_l {\sigma_l}^2$. 
This upper bound is reached when 
$\psi + \alpha_l - \alpha_k = 0$,
as well as when
$\bar{l}'=0$
or 
$\bar{k}'=0$. 
This optimization can be achieved by an adjustment of the
the rotation phases to the value 
$\beta = \pi - 2(\alpha_k - \alpha_l)$. 
However, this requires one to know the difference between the phases
$\alpha_k$ and $\alpha_l$, which is not generally available
for an arbitrary initial amplitude distribution. 

The optimal case of 
$P_{\max}=1$ 
can be
obtained by using the predefined state 
$\ket{s}$ 
and applying on it the operator $U$, to generate the
initial amplitude distribution. 
In this case the initial unmarked state variance is 
$\sigma_l = 0$, 
the weighted initial averages are 
$\bar{k}'(0) = 1$ 
and 
$\bar{l}'(0) = 1$,
and the phases $\alpha_k = \alpha_l = 0$. 
Thus, executing 
the generalized Grover iterations using the same unitary operator $U$, 
the predefined state $\ket{s}$ as the initial state 
and a rotation phase of $\beta = \gamma = \pi$ enables one to measure a marked 
state with the optimal probability $P_{\max} = 1$. 
The original Grover 
algorithm is a special case in which $U$ is the Hadamard operator and 
the predefined state $\ket{s} =  \left| 0 \ldots 00 \right\rangle$. 
 
Consider an arbitrary initial distribution of $r$ marked and $N-r$ unmarked 
states, with known averages 
$\bar{k}(0)$ 
and 
$\bar{l}(0)$ 
respectively. 
The probability $P(t)$ that a measurement at time $t$ will yield a marked state
is a sinusoidal function, given by
Eq.~(\ref{eq:Pt}). The highest value of $P(t)$ is obtained at time $T$, for
which the argument of the cosine function satisfies
$2(\omega T + \phi) = \pi$.
Thus, the number of iterations $T$ which gives rise to the highest 
probability of finding a marked state upon measurement is 

\begin{equation}
T = \frac{ \pi - 2 \phi }{ 2 \omega },  
\label{eq:T}
\end{equation} 

\noindent
where the angular frequency is

\begin{equation}
\omega = 2 \sin {\beta \over 2} \sqrt{W_k} + O(W_k^{3/2}).
\label{eq:omegaBeqG}
\end{equation} 

An interesting case is the one in which the average and variance of the 
initial amplitude distribution are {\it not} known, 
but different runs of the algorithm use initial amplitudes drawn from the 
same distribution.
Naively, one could pick a random number of iterations $T_r$ and thus find 
a marked state with probability $P(T_r)$.
Correspondingly, the expected number of repetitions of the entire algorithm 
using the same $T_r$ would be $1/P(T_r)$ until a marked state is found.   
However, $P(T_r)$ could be very small. 
A better strategy is now shown. 
>From Eqs.~(\ref{eq:cosw}) 
and (\ref{eq:Pt})
it follows that the period of oscillation of $P(t)$ depends only on 
the unitary operator $U$, the predefined state
$\ket{s}$
as well as on the rotation phase 
$\beta=\gamma$
used in the algorithm. 
The phase $\phi$ depends on
the initial amplitude distribution
and is thus unknown.
Consider the case where one runs the algorithm twice, taking measurements at 
times $T_1$ and $T_2$ respectively, where $T_2-T_1 = \pi/(2 \omega)$.
>From Eq. (\ref{eq:Pt}) it is clear that in one of the two measurements 
the cosine expression will be negative so that $P(T) \geq P_{av}$. 
Since the probability $P(t)$ must be non-negative at any time, 
$P_{av} \geq \Delta P$.
Since 
$P_{\max} = P_{av} + \Delta P$, we also find that
$P_{av} \geq P_{\max}/2$.
Therefore, in one of the two measurements one obtains
$P(T) \geq P_{max}/2$.
In this case one needs twice as many repetitions to obtain at least
half the success probability compared to the case when the optimal 
measurement time is known.
The slowdown is thus at most by a factor of 4.

\section{Summary}

In this paper we have generalized the
recursion equation analysis of Grover's quantum search 
algorithm presented 
in Ref.
\cite{BihamE99}.
We applied it to the 
large class of Grover's type algorithms in which the Hadamard 
transform is replaced by any unitary transformation 
and the phase inversion is replaced by a rotation by an arbitrary
angle.
We derived recursion equations for
the time evolution of the amplitudes of the marked and 
unmarked states, 
for any initial complex amplitude distribution. 
These equations were solved {\it exactly}. 
>From the solution we obtained
an expression for
the optimal number of iterations $T$ after which 
the probability of finding a marked state upon measurement is the highest. 
The value of this probability, $P_{\max}$, was also obtained.  
Both $T$ and $P_{\max}$ 
are found to depend on the averages
and variances of the initial amplitude distributions 
of the marked and unmarked states,
but not on higher moments.
This is due to the linearity of the transformations and the large number
of conserved quantities, particularly the (weighted) variances of the distributions
of the amplitudes of the marked and unmarked states.
The time $T$ and the probability $P_{\max}$ also
depend on the unitary operator
$U$ which is used in the algorithm, the predefined state
$\ket{s}$,
the angle $\beta$ by which its phase is rotated as well as
the angle $\gamma$ by which the phases of the marked states
are rotated. 
Moreover, it was found that in order for the algorithm to apply
the two rotation angles must be equal, namely
$\beta=\gamma$.

\section{Acknowledgements}

We thank Dan Kenigsberg for helpful discussions and comments.
This work was initiated during the Elsag-Bailey --
I.S.I. Foundation research meeting on quantum computation in 1998. 
It was supported by the EU fifth framework program grant 
number IST-1999-11234.
	
\appendix  
\section{}

In this Appendix we obtain upper bounds on
$|\bar{k}'(0)|$
and
$|\bar{l}'(0)|$
in the initial amplitude distribution.
According to  
Eqs.~(\ref{vars_k:eq}), 
and 
(\ref{eq:kaverage}) 
the initial  
distribution weighted average of the marked states is:  

\begin{equation}
\bar{k}'(0) = { {\sum_{i\in M} {\eta_i}^{*}  k_i(0)} \over W_k }.
\label{eq:kbar0}
\end{equation}

\noindent
>From normalization it is clear that
$|k_i(0)| \leq 1$ 
for any marked state. 
Therefore

\begin{eqnarray}
|\bar{k}'(0)| &\leq&  { {\sum_{i\in M} |\eta_i| | k_i(0)|} \over W_k } 
\leq  { {\sum_{i\in M} |\eta_i|} \over W_k } \leq  
{{r  |\eta|}  \over W_k} 
\label{eq:kbar0ineq}
\end{eqnarray}

\noindent
where $r$ is the number of marked states and $|\eta| = \max_{i \in M} 
|\eta_i|$. 
Since 
$|\eta|^2 \leq  \sum_{i\in M} |\eta_i|^2 = W_k  $ 
one obtains

\begin{eqnarray}
|\bar{k}'(0)| &\leq& {r {W_k}^{\frac{1}{2}} \over W_k} = 
{r  {W_k}^{-\frac{1}{2}}}. 
\label{eq:kbar0O}
\end{eqnarray}

\noindent
Therefore, typical initial amplitude distributions in large search
problems satisfy

\begin{equation}
|\bar{k}'(0)| = O( {W_k}^{-\frac{1}{2}} ). 
\label{eq:Okbar}
\end{equation}

\noindent
According to  Eqs.~(\ref{vars_l:eq}) 
and 
(\ref{eq:laverage}) 
the 
weighted average of the
initial distribution 
$l_i'(0)$
satisfies

\begin{eqnarray}
\bar{l}'(0) &=& { {\sum_{i\in \overline{M}} {\eta_i}^{*}  l_i(0)} \over W_l } 
= \sum_{i\in \overline{M}} {{\eta_i}^{*}  l_i(0)} + O(W_k). 
\label{eq:lbar0}
\end{eqnarray}

\noindent
Using Eq. (\ref{ket_gt}) for the initial state of the system $\ket{g(0)}$,
the expression for 
$\bar{k}'(0)$ 
[given in Eq. (\ref{eq:kbar0})] 
and 
${\eta_i}^{*} = \bra{s} {U}^{*} \ket{i}$, 
one can bring 
Eq. (\ref{eq:lbar0}) to the form: 

\begin{equation} 
\bar{l}'(0) = \bra{s} {U}^{*} \ket{g(0)} - W_k \bar{k}'(0) + O(W_k). 
\label{eq:lbar01}
\end{equation}

\noindent
Using 
Eq.~(\ref{eq:Okbar})
one obtains

\begin{equation} 
\bar{l}'(0) = \bra{s} {U}^{*} \ket{g(0)} +  
O({W_k}^{\frac{1}{2}}). 
\label{eq:lbar02}
\end{equation}

Since 
$\ket{g(0)}$ 
is normalized 
and ${U}^{*}$ is unitary, 
$| \bra{s} {U}^{*} \ket{g(0)} | < 1$. 
Therefore,

\begin{equation} 
|\bar{l}'(0)| < 1 +  O({W_k}^{\frac{1}{2}}),
\label{eq:lbar03}
\end{equation}

\noindent
namely

\begin{equation}
|\bar{l}'(0)| = O(1). 
\label{eq:Olbar}
\end{equation}

\end{multicols}


\begin{thebibliography}{10}

\bibitem{Grover96}
{L.K. Grover},  in {\em {Proceedings of the Twenty-Eighth Annual Symposium on
  the Theory of Computing }} ({ACM Press}, {New York}, {1996}), p.\ 212.

\bibitem{Grover97a}
{L.K. Grover}, Phys. Rev. Lett. {\bf 79},  325  (1997).

\bibitem{Stinson95}
{D.R. Stinson}, {\em {Cryptography: Theory and Practice}} ({C.R.C Press}, {},
  1995).

\bibitem{Garey:book}
{M.R. Garey and D.S. Johnson}, {\em {Computers and Intractability: A Guide to
  the Theory of NP-Completeness}} ({W.H. Freeman}, {New York}, 1979).

\bibitem{Zalka99}
{C. Zalka}, Phys. Rev. A {\bf 60},  2746  (1999).

\bibitem{Bennett97}
{C.H. Bennett, E. Bernstein, G. Brassard and U. Vazirani}, {SIAM J. Comput.}
  {\bf 26},  1510  (1997), {e-print quant-ph/9701001}.

\bibitem{Durr96}
{C. Durr and P. H{\o}yer}, {e-print quant-ph/9607014}.

\bibitem{Grover97b}
{L.K. Grover}, {e-print quant-ph/9607024}.

\bibitem{Grover97c}
{L.K. Grover}, {e-print quant-ph/9704012}.

\bibitem{Terhal98}
{B.M. Terhal and J.A. Smolin}, Phys. Rev. A {\bf 58},  1822  (1998).

\bibitem{Brassard98}
{G. Brassard, P. H{\o}yer and A. Tapp}, {in Automata, Languages and
  Programming, Vol. 1443, (Springer Verlag, Berlin, 1998), p. 820, e-print
  quant-ph/9705082}.

\bibitem{Cerf00}
{N.J. Cerf, L.K. Grover and C.P. Williams}, Phys. Rev. A {\bf 61},  2303
  (2000).

\bibitem{Grover00}
{L.K. Grover}, Phys. Rev. Lett. {\bf 85},  1334  (2000).

\bibitem{Carlini99}
{A. Carlini and A. Hosoya}, {e-print quant-ph/9909089}.

\bibitem{Chuang98b}
{I.L. Chuang, N. Gershenfeld and M. Kubinec}, Phys. Rev. Lett. {\bf 80},  3408
  (1998).

\bibitem{Jones98}
{J.A. Jones, M. Mosca and R.H. Hansen}, Nature {\bf 393},  344  (1998).

\bibitem{Kwiat00}
{P.G. Kwiat, J.R. Mitchell, P.D.D. Schwindt and A.G. White}, J. Mod. Optics {\bf
  47},  257  (2000).

\bibitem{Boyer96}
{M. Boyer, G. Brassard, P. H{\o}yer and A. Tapp},  in {\em {Proceedings of the
  fourth workshop on Physics and Computation }}, edited by {T. Toffoli, M.
  Biafore and J. Leao} ({New England Complex Systems Institute}, {Boston},
  {1996}), p.\ 36.

\bibitem{Grover98a}
{L.K. Grover}, Phys. Rev. Lett. {\bf 80},  4329  (1998).

\bibitem{Gingrich00}
{R.M. Gingrich, C.P. Williams and N.J. Cerf}, Phys. Rev. A {\bf 61},  2313
  (2000).

\bibitem{Jozsa99}
{R. Jozsa}, {e-print, quant-ph/9901021}.

\bibitem{Long99a}
{G.L. Long, W.L. Zhang, Y.S. Li and L. Nui}, Commun. Theor. Phys. {\bf 32},
  335  (1999).

\bibitem{Biron97}
{D. Biron, O. Biham, E. Biham, M. Grassl and D.A. Lidar}, {Generalized Grover
  search algorithms for arbitrary initial amplitude distribution, Proceedings
  of the 1st NASA International Conference on Quantum Computing and Quantum
  Communications, Lecture Notes in Computer Science (Springer-Verlag, 1998);
  e-print quant-ph/9801066}.

\bibitem{BihamE99}
{E. Biham, O. Biham, D. Biron, M. Grassl and D. Lidar }, Phys. Rev. A {\bf 60},
   2742  (1999).

\bibitem{Long99b}
{G.L. Long, C.C. Tu, Y.S. Li, W.L. Zhang and H.Y. Yan}, {e-print
  quant-ph/9911004}.

\end{thebibliography}
\end{document}